\documentclass[12pt]{article}

\usepackage{amsmath}  
\usepackage{amssymb}  
\usepackage{bm}
\usepackage{stmaryrd}  

\usepackage{xcolor}
\usepackage{hyperref}  
\hypersetup{%
    colorlinks=false,
    linkbordercolor=white,
    urlbordercolor=cyan,  
    pdfborderstyle={/S/U/W 1}  
}

\usepackage{graphicx}
\usepackage{authblk}

\newcommand{\boldvec}[1]{{{\mathbf{#1}}}}
\providecommand{\like}{{\mathcal{L}}}
\newcommand{\vD}{\boldvec{D}}
\newcommand{\vM}{\boldvec{M}}
\providecommand{\paramvec}{{\boldvec{\theta}}}

\newcommand{\sparam}{{\cal S}}
\newcommand{\oparam}{{\cal O}}
\newcommand{\intens}{\mu}

\newcommand{\effic}{\eta}

\newcommand{\dif}{\textrm{d}}  

\begin{document}

\title{Multilevel and hierarchical Bayesian modeling\\ of cosmic populations%
\thanks{The main text of this paper is an update of ``Bayesian multilevel modelling of cosmological populations,'' which appeared in \textsl{Bayesian Methods in Cosmology} (ed.\ by M.~P.~Hobson, A.~H.~Jaffe, A.~R.~Liddle, P.~Mukerherjee, D.~Parkinson), Cambridge University Press, pp.~245--264 (2010)
(\href{https://doi.org/10.1017/CBO9780511802461.012}{DOI}).
The Appendix is drawn from a monograph in preparation; the authors would be grateful for comments on the new material.}}
\author[1]{Thomas J. Loredo}
\author[2]{Martin A. Hendry}
\affil[1]{Cornell Center for Astrophysics and Planetary Science, Cornell University}
\affil[2]{School of Physics \& Astronomy, University of Glasgow}
\setcounter{Maxaffil}{0}
\renewcommand\Affilfont{\itshape\small}

\date{}
\maketitle


\noindent
\textbf{Abstract:}
Demographic studies of cosmic populations must contend with measurement errors and selection effects. We survey some of the key ideas astronomers have developed to deal with these complications, in the context of galaxy surveys and the literature on corrections for Malmquist and Eddington bias. From the perspective of modern statistics, such corrections arise naturally in the context of multilevel models, particularly in Bayesian treatments of such models: hierarchical Bayesian models. We survey some key lessons from hierarchical Bayesian modeling, including shrinkage estimation, which is closely related to traditional corrections devised by astronomers. We describe a framework for hierarchical Bayesian modeling of cosmic populations, tailored to features of astronomical surveys that are not typical of surveys in other disciplines. This thinned latent marked point process framework accounts for the tie between selection (detection) and measurement in astronomical surveys, treating selection and measurement error effects in a self-consistent manner.

\bigskip

\section{Introduction}
\label{LH-sec:intro}

\begin{figure}[t]
\centerline{\includegraphics[width=.95\textwidth]{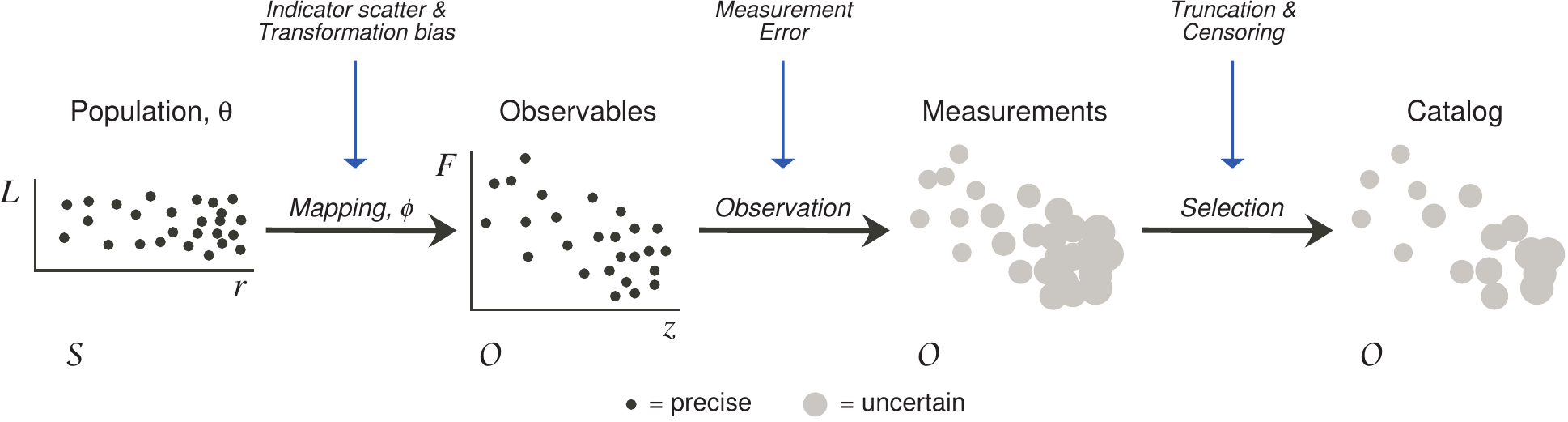}}
\caption{\small Schematic depiction of the survey process.
A population with a distribution of
source or object properties, $\sparam$ (e.g, luminosity and distance), implies,
via a mapping $\phi$, a distribution of observables, $\oparam$ (e.g., flux
and redshift).  Observation introduces measurement error; selection 
criteria thin, truncate, or censor the cataloged population.}
\label{LH-SurveyCartoon}
\end{figure}

Surveying the Universe is the ultimate remote sensing problem.
Inferring the intrinsic properties of the galaxy population, via
analysis of survey-generated catalogs, is a major challenge for
twenty-first century cosmology, but this challenge must be met
without any prospect of measuring these properties {\em in situ\/}.
Thus, for example, our knowledge of the intrinsic luminosity and spatial
distribution of galaxies is filtered by imperfect distance
information and by observational selection effects---issues which
have come to be known generically in the literature as ``Malmquist
bias.''\footnote{Although ``Malmquist'' is the most prevalent appellation,
the literature also uses other terms---including
``Eddington-Malmquist'' and ``Lutz-Kelker''---to denote biases
arising in astronomical surveys from distance indicator scatter and
observational selection. There is also an unfortunate
history in the cosmology literature of the same term being used to
mean substantially different things by different authors. For a more
detailed account of the meaning, use, and abuse of bias terminology
see, e.g., Hendry and Simmons (1995), Strauss and Willick (1995) and
Teerikorpi (1997).}
Figure~\ref{LH-SurveyCartoon} shows schematically 
how such effects may distort our inferences about the underlying population
since in general these must be derived from a noisy, sparse and truncated
sample of galaxies.

There is a long (and mostly honourable!) tradition in the
astronomical literature of attempts to cast such remote surveying
problems within a rigorous statistical framework. Indeed, it is
interesting to note that seminal examples from the early twentieth
century---e.g., Eddington (1913, 1940); Malmquist (1920, 1922)---display, at
least with hindsight, hints of a Bayesian formulation long before
the recent renaissance of Bayesian methods in astronomy.
Space does not permit us to review in detail that
early literature, nor many of the more recent papers which
evolved from it.  A more thorough discussion of the early literature on
statistical analysis of survey data can be found in, e.g., Hendry
and Simmons (1995), Strauss and Willick (1995), Teerikorpi (1997),
and Loredo (2007).

The analysis of survey catalogs can have a number of scientific
goals. For example, the objective may be to compare the underlying
galaxy luminosity distribution with the predictions of different
galaxy formation models. In this case the galaxy distances (which we
must infer as an intermediate step towards estimating their
luminosities) are, in statistical parlance, nuisance parameters. On the
other hand, the goal may be to infer the distances of the surveyed
galaxies, in order to test models of galaxy clustering and/or
constrain parameters of the underlying cosmology. In this case it is
the inferred galaxy luminosities (or other properties) which may be thought of as nuisance parameters.

In the literature of the past 25 years, the second case---that of
inferring galaxy distances---has proven to be fertile territory for
the development and application of Bayesian methods. This is
particularly true with regard to (largely) redshift-independent distance
indicators, i.e., indicators whose behaviour is independent of the
underlying cosmological model (and which may thus be
straightforwardly used to constrain parameters of that model).  A
likely reason for this is that redshift-independent distance
indicators suffer from large intrinsic scatter, with distance
uncertainties to individual galaxies typically in the range 5\% to
30\%.  A consequence is that care must be taken when incorporating
prior distance information, since in this setting final inferences
can be significantly influenced by the prior, a point we discuss
further in Section~\ref{LH-sec:indic}.

Another hallmark of the literature on extracting information from
galaxy surveys has been the recognition by some authors (e.g.,
Hendry and Simmons 1995; Loredo 2007) that the task of identifying
``optimal'' (e.g., in the sense of unbiased and/or minimum mean square
error) estimators of galaxy distance and luminosity will generally
not have a unique solution, but will depend of the context in which
the inferred galaxy distances or luminosities are to be used.

In Section \eqref{LH-sec:multi} we discuss how multilevel
models, and particularly hierarchical Bayes\-ian models provide a natural, powerful framework in which to formulate
and implement optimal analyses of surveys, incorporating prior
information and carefully accounting for selection effects and
source or object uncertainties. To motivate this approach, and to establish
context, in Section \eqref{LH-sec:indic} we survey key Bayesian
elements of recent methodology for analysis of
survey data, focusing on use of redshift-independent distance
indicators, and highlighting examples which have previously hinted
at a multilevel approach.

\section{Galaxy distance indicators}
\label{LH-sec:indic}

While astrometric space missions (e.g. \emph{Gaia}) offer some
prospect of applying trigonometric methods over nearby cosmological scales,
in large part the measurement of galaxy distances relies on more
astrophysical, and therefore less precise, methods. Perhaps most obvious
among these methods is the cosmological redshift.  To derive a
distance measure from it requires the use of a cosmological
model.  But if one's goal
is to use galaxy distance estimates to probe the parameters
of the cosmological model, the safest path is to identify galaxy distance
indicators whose properties are largely independent of redshift.

Almost all redshift-independent distance indicators are really
\emph{luminosity} or \emph{size} indicators:  one uses the indicator to estimate
the intrinsic luminosity or size of a galaxy (or some
object therein); combined with a measurement of its apparent brightness or
apparent size, one may estimate distance via the
inverse-square law or the angle-distance relation (or their cosmological
generalisations). For
simplicity we consider here only the case of luminosity indicators,
although very similar statistical considerations apply to other
types of indicator.  We can estimate the luminosity either by
assuming a constant, fiducial value (the so-called `standard candle'
assumption) or, better, by exploiting correlations between
luminosity and some other intrinsic, but directly measurable,
physical characteristic(s) of the source. (The latter approach is
often referred to as a `standardisable candle.') Examples of these
correlations include: the Tully-Fisher relation for spiral galaxies;
the period-luminosity relation for Cepheid variables and the
luminosity--light curve shape relation for type Ia supernovae. The
correlations may be motivated by theory, but they must be calibrated
empirically, e.g., using nearby galaxies at known distances (and hence
of known luminosity). Their scatter, reflecting the intrinsic spread in
luminosity of the objects, renders distance indicators susceptible
to observational selection biases since one cannot observe
arbitrarily faint objects to arbitrary distances.

Since the late 1980s several authors have investigated the
statistical properties of redshift-independent galaxy distance
indicators, with the goal of placing their use in cosmology on a
more rigorous statistical footing. A significant step forward in
this regard came in the early 1990s with the work of Jeffrey Willick
(1994), which brought much needed clarity to the discussion by
explictly making a distinction between the tasks of \emph{calibrating} a
distance indicator and applying it to a galaxy survey to \emph{estimate}
distances. We now briefly summarise the formalism
presented in Willick (1994) and adopted in subsequent papers.

\subsection{The calibration problem}

Our starting point is the joint probability distribution for a
single galaxy's distance, $r$, apparent magnitude, $m$, and some
third observable correlated with luminosity which, following
Willick, we denote by $\eta$ and refer to as the `line width'
parameter.  As a concrete example, consider the Tully-Fisher
relation, for which we expect the intrinsic relation between
absolute magnitude and $\eta$ to be linear, i.e. $M = a \eta + b$,
where the coefficients $a$ and $b$ must be calibrated empirically.
As noted earlier, an analysis may have various goals: $a$ and $b$
may simply be nuisance parameters, necessary to estimate galaxy
distances; alternatively, they may be important target parameters in
their own right.

Thus, for the joint probability density function (PDF) describing the properties of galaxies within a particular survey catalog, we have
\begin{equation}
p(r,m,\eta) \propto r^2 n(r) \, S(m,\eta) \, \psi(m | \eta) \,
\phi(\eta),
\label{LH-gal-distn}
\end{equation}
where $n(r)$ and $\phi(\eta)$ denote the marginal distributions of
distance and line width respectively for the galaxy population (we will assume $r$ and $\eta$ are independent),
$\psi(m | \eta)$ denotes the conditional distribution of apparent
magnitude at a given line width (which depends on unknown parameters,
e.g., $a$ and $b$), and $S(m,\eta)$ denotes the
observational selection effects (a detection probability, assumed here, for simplicity, not to depend on distance or direction).

Consider first the calibration of the distance indicator, which
might reasonably be carried out, e.g., using a galaxy cluster, so
the set of calibrators are effectively all at the same
distance.\footnote{Willick also considers the calibrators at a range
of true distances; this case lends itself well to a Bayesian
multilevel formulation, as we discuss in Section \eqref{LH-sec:multi}}
In this case it is natural to work with the conditional distribution
of $m$ at given $\eta$ and $r$ (i.e., the result from using
equation~\eqref{LH-gal-distn} as a prior in Bayes' theorem, with a
``likelihood'' corresponding to precise measurement of $\eta$ and
$r$).  Then the indicator coefficients $a$ and $b$ can be
interpreted as the slope and zero-point of a linear regression of
absolute magnitude on $\eta$---this case is known as the `direct'
indicator relation. Thus
\begin{equation}
P(m | \eta, r) = \frac{ S(m, \eta) \, \psi(m | \eta)}{ \int_{-
\infty}^{\infty} \, S(m, \eta) \, \psi(m | \eta) dm}.
\end{equation}
Notice that, as expected, the marginal distributions of distance and
line width drop out. The presence of the observational selection
effects will bias the determination of the indicator coefficients
$a$ and $b$ obtained via simple linear regression; however, Willick
(1994) proposed an iterative scheme to overcome this problem and
showed that it works well for realistic mock galaxy data.

A popular variant on the above approach is to use the so-called
`inverse' relation, i.e. (in Willick's notation) $\eta^0(M) = a' M +
b'$, where the inverse coefficients $a'$ and $b'$ again must be
determined empirically (and again may be regarded either as nuisance
parameters or target parameters).  This relation is most directly
expressed by
the conditional distribution of $\eta$ at a given $r$ and $m$ (corresponding
to given $M$, since we are assuming all the calibrators
lie at the same distance), namely
\begin{equation}
p(\eta | m, r) = \frac{ S(m, \eta) \, \Psi(\eta | m)}{ \int_{-
\infty}^{\infty} \, S(m,\eta) \, \psi(\eta | m) d\eta},
\end{equation}
where $\Psi$ denotes the conditional distribution of line width at
given apparent magnitude (i.e., the reverse of the conditioning in $\psi(m|\eta)$). In this case explicit dependence on the
galaxy luminosity function (the distribution for $M$) drops out of our expression (because conditioning on $(m,r)$ amounts to conditioning on $M$). Moreover,
one can see that {\em if\/} the selection effects depend only on
apparent magnitude and not on line width, then a straightforward
linear regression of $\eta$ on $M$ {\em will\/} yield unbiased
estimates of the indicator coefficients $a'$ and $b'$. This
appealing property of the inverse indicator relation had been
recognised in principle much earlier by Schechter (1980) and was
also placed on a rigorous statistical footing around the same time
as Willick by Hendry and Simmons (1994). However, the successful
calibration of a galaxy distance indicator is only the first part of
the story.

\subsection{The estimation problem}
\label{LH-sec:estim}

Suppose one has used the relations above to accurately and precisely
calibrate a distance indicator (e.g., $a$ and $b$ are now precisely known).
Now we seek to use the indicator in settings where there is no direct
measurement of $r$; we must infer $r$ from measurements of $m$ and $\eta$.
We can calculate a predicted galaxy distance, $d$, in the obvious way by combining
the observed apparent magnitude of the galaxy with its estimated
absolute magnitude inferred (via our indicator relation) from its
observed line width. Moreover, since $d = d(m,\eta)$, it is
straightforward to compute the joint distribution, $p(r,d)$, of true
and estimated galaxy distance, and further to determine the
conditional distribution of $r$ given $d$. For the direct indicator
we obtain
\begin{equation}
p(r|d) = \frac{ r^2 n(r) \, \exp \left ( - \frac{ [ \ln r/d ]^2}{2
\Delta^2} \right )}{ \int_0^\infty r^2 n(r) \, \exp \left ( - \frac{
[ \ln r/d ]^2}{2 \Delta^2} \right ) dr},
\label{eq:dirprob}
\end{equation}
where $\Delta$ is a constant describing the scatter in the direct indicator relation, i.e. the dispersion of the conditional distribution of absolute magnitude at given line width (here assumed Gaussian).
For the inverse indicator, on the other hand, we obtain
\begin{equation}
p(r|d) = \frac{ r^2 n(r) s(r) \, \exp \left ( - \frac{ [ \ln r/d
]^2}{2 \Delta^2} \right )}{ \int_0^\infty r^2 n(r) s(r) \, \exp
\left ( - \frac{ [ \ln r/d ]^2}{2 \Delta^2} \right ) dr},
\label{eq:invprob}
\end{equation}
where $s(r)$ is an integral over the galaxy luminosity function
weighted by the selection effects, and expresses the probability
that a galaxy at true distance $r$ would be observable in the
survey. This term is often referred to as the {\em selection
function\/} for $r$.

The interpretation of equations (\eqref{eq:dirprob}) and
(\eqref{eq:invprob}) within the framework of Bayesian inference is
clear. We can think of $p(r|d)$ as representing the posterior
distribution of true distance $r$, given some observed data $d$
(i.e. the estimated distance, from our indicator). Moreover the {\em
difference\/} between the two expressions can then be interpreted in
terms of the adoption of different prior information for $r$: for
the direct indicator the prior information is the true distance
distribution $n(r)$, while for the inverse relation the prior is the
product of $n(r)$ and the selection function.

The classical Malmquist bias is manifest when we take the
conditional expectation of $r$ given $d$, using equations
(\eqref{eq:dirprob}) and (\eqref{eq:invprob}). For both direct and
inverse indicators we find that in general $E(r|d) \neq d$. However
we can correct our `raw' distance indicator $d$, defining $d_{\rm
corr}$ which satisfies
\begin{equation}
E(r|d_{\rm corr}) = d_{\rm corr},
\label{eq:malmcorr}
\end{equation}
with the correction term referred to as a ``Malmquist correction.''
Note, however, that the Malmquist correction depends explicitly on
the true distance distribution $n(r)$, which in general will be
unknown. Lynden-Bell et al. (1988) computed homogeneous Malmquist
corrections, assuming that the underlying spatial distribution of
galaxies is uniform, in which case
\begin{equation}
E(r|d) \;=\; d \, \exp \left ( \frac{7}{2} \Delta^2 \right ) \;
\simeq \; d \, \left ( 1 + \frac{7}{2} \Delta^2 \right ) \; \equiv
\; d_{\rm corr}.
\label{eq:hmc}
\end{equation}
We should not be surprised that the correction is always positive in
this case; since we are assuming homogeneity we are saying that the
distance indicator scatter is more likely to scatter galaxies {\em
downwards\/} from greater true distances, simply because there are
more galaxies at larger $r$ due to the rapid growth of the volume
element with $r$.  Note, however, that for the inverse indicator the
assumption of a uniform prior is not appropriate: even if $n(r)$
were constant, the selection function $s(r)$ clearly will not be.

The more realistic case is, of course, where the intrinsic
distribution of distance is {\em not\/} uniform. In this case the
adoption of a suitable prior for $n(r)$ leads to a so-called
\emph{inhomogeneous Malmquist correction}. For the direct indicator
the source of the prior information could be, for example, the
underlying density field of galaxies reconstructed from an external
source, e.g., an all-sky redshift survey (c.f.\ Hudson 1994; Strauss
and Willick 1995; Freudling et al.\ 1995; Ergogdu et al.\ 2006). The
Malmquist corrections will only be valid in this case, however,
provided that the external galaxy survey traces the same underlying
population as the galaxies to which the distance indicator is being
applied.

In an important paper, Landy and Szalay (1992) proposed an
interesting alternative approach, whereby the marginal distribution
of raw distances might provide a suitable estimate of the prior true
distance distribution. Crucially, this method should {\em not\/} be
applied using the direct indicator since the marginal distribution
of raw distances provides a poor estimate of $n(r)$. On the other
hand, it {\em does\/} provide a reasonable proxy for the
distribution of true distances for `observable' galaxies---i.e. the
product of $n(r)$ and $s(r)$. Thus, it is probably well suited to
use with the inverse indicator.

The Landy and Szalay approach, although invoking implicit approximations, has
several attractive features. It offers a method of defining
inhomogeneous Malmquist corrections that adapts to spatial inhomogeneity,
without requiring external assumptions or prior information
about $n(r)$ from other galaxy surveys. Indeed it appears to be an
approximation to a hierarchical Bayesian procedure, as we discuss
further in Section~\ref{LH-sec:multi}.

\subsection{Applications of galaxy distance indicators}
\label{LH-sec:app}

Why might we regard Malmquist-corrected distance indicators, which
satisfy equation (\eqref{eq:malmcorr}), as optimal estimators in the
first place? The answer lies largely in the uses to which they have
been put.
In the late 1980s redshift-independent distance
indicators began to be used to measure galaxy {\em peculiar
velocities\/}---the motions, over and above the Hubble expansion,
induced by the net gravitational attraction of the matter
distribution around them. Methods of analysing peculiar velocities
generally involve first binning and grouping galaxies together based
on their {\em estimated\/} distance. By requiring that on average
the true distance of each galaxy be equal to its estimated distance,
one aims to ensure that on average the correct radial peculiar
velocity will be ascribed to each galaxy's apparent position.

In the 1990s astronomers developed a number of sophisticated methods
to compare observed and predicted galaxy peculiar velocities, the
latter the result of reconstructing the density and peculiar
velocity field from position and redshift data from an all-sky
redshift survey.  This reconstruction requires a model for
\emph{galaxy biasing}---i.e., a description of how the distribution
of luminous galaxies and dark matter are related. By comparing
observed and predicted peculiar velocities one can constrain
parameters of the galaxy biasing model.

From a Bayesian perspective probably the most notable of these
comparison methods was VELMOD (Willick and Strauss 1998). This
assumed a simple linear relation between the galaxy and matter
density fields and computed a posterior distribution for the linear
bias parameter, marginalized over the nuisance parameters of the
distance indicator relation. In its explicit modeling of the galaxy
distance uncertainties, en route to estimating the linear bias
parameter, VELMOD shares features with the hierarchical Bayesian
approach which we now describe.

\section{Multilevel, hierarchical Bayesian models}
\label{LH-sec:multi}

The issues motivating the astronomical developments just surveyed are hardly
unique to astronomy.  Statisticians have addressed similar issues in
applications spanning many disciplines.  Although none of the resulting methods
is an ``exact fit'' to an astronomical survey problem, the body of literature
offers numerous insights that are inspiring significant advances in
Bayesian methodology for astronomical surveys.

A recurring theme of much of the relevant literature is the use of
\emph{hierarchical Bayesian models} (HBMs).
These are a subclass of \emph{probabilistic graphical models}, and represent a Bayesian treatment of \emph{multilevel models} (MLMs, terminology used in both Bayesian and frequentist literature).
These terms cover a rich framework
that underlies several important statistical innovations of the latter
20th century, including empirical Bayes methods,
random effects and latent variable models, shrinkage estimation, and ridge
regression.  MLMs start with a \emph{lower level} probability model for
the data, given \emph{latent parameters} specifying the properties of members of the surveyed population (e.g., objects or sources).
This level typically describes noise and other aspects of the individual member measurement process.
The \emph{upper level} assigns a shared prior distribution to the
lower-level parameters (e.g., a population-level distribution for object
properties); this
distribution may itself have unknown parameters, dubbed
\emph{hyperparameters}.  The upper level leads to probabilistic dependence
among the lower-level latent parameters that implements a pooling of information that
can improve the accuracy of inferences; one says the estimates ``borrow
strength'' from each other.  Other levels may be added, e.g., to describe
relationships between groups of objects.

We here focus on Bayesian treatment of MLMs, though multilevel modeling is an
area where there has been significant cross-fertilization between Bayesian
and frequentist approaches.  We begin by describing a very simple MLM---the
\emph{normal--normal} MLM---highlighting a feature of MLM point
estimates---shrinkage---that
has connections to classic astronomical approaches for correcting for survey
biases.  We use this as a stepping stone to a more thoroughgoing Bayesian
approach that moves beyond point estimates and corrections.
Until recently this
approach has been implemented only in fairly simple astronomical settings; we
end by highlighting directions for future research.

\subsection{Adjusting member estimates: shrinkage}

Suppose we have survey data for a population of objects that we will model as
having a log-normal luminosity function, so the population distribution of
absolute magnitude, $M$, is a normal distribution with location $M_0$
and scale (standard deviation) $\tau$ (these are the hyperparameters).
As a simple starting point, we will suppose $\tau$ is known ($\tau = 0.5$~mag,
say), and we denote the population distribution by $f(M|M_0)$.   For now,
we also assume there are no selection effects.  At the population level, our
goal is to infer $M_0$.

The survey produces data, $D_i$, for each object; we will suppose
these lead to independent, Gaussian-shaped likelihood functions for
each object's unknown true absolute magnitude $M_i$, with maximum
likelihood estimates (MLEs) $\hat M_i$ and uncertainties (standard
deviations) $\sigma_i$.  We denote these \emph{member} or \emph{object likelihood functions} as
\begin{equation}
\ell_i(M_i) \equiv p(D_i|M_i) = N(M_i|\hat M_i,\sigma^2_i),
\label{mlike}
\end{equation}
with $N(\cdot|\mu,\sigma^2)$ the normal distribution with mean $\mu$ and
variance $\sigma^2$.  The $M_i$ are the lower level parameters.  The
survey catalog consists of a table of the $\hat M_i$ estimates and
their uncertainties, understood here as descriptions of member likelihood functions.  For simplicity, we assume equal uncertainties,
$\sigma_i = \sigma = 0.3$~mag.

Let $\vD \equiv \{D_i\}$ and $\vM \equiv \{M_i\}$ denote the
collections of data and object parameters, respectively.
The likelihood function for the population parameter, $M_0$, and all of the latent member parameters, $\vM$, is
\begin{equation}
\like(M_0,\vM) \equiv p(\vD|M_0,\vM) = p(\vD|\vM)
=\prod_i\ell_i(M_i).
\end{equation}
Note that it does not depend on $M_0$ because
the probabilities for the source data, $D_i$, are fully determined
(and independent) if $\vM$ is specified.  Thus the joint MLEs for
the source parameters are just the independent MLEs: $\hat \vM =
\{\hat M_i\}$. But in a Bayesian calculation, estimates are
determined by the posterior, not the likelihood.  If $M_0$ is known,
the joint posterior for the object parameters, conditional on $M_0$,
is given by
\begin{equation}
\pi(\vM|\vD,M_0) \propto \prod_i f(M_i)\, \ell_i(M_i),
\label{LH-cond-post}
\end{equation}
with the population density appearing as a prior factor for each source.
Point estimates may be found from this conditional posterior, e.g., by finding
the mode or posterior mean for $\vM$.
It is important to note that if we increase the amount of survey data
by increasing $N$, such Bayesian estimates will \emph{not} converge to
the MLEs, because additional prior factors enter with each new source.
That is, we are not in the common, simpler setting of a fixed number
of parameters, with additional data providing likelihood factors that eventually
overwhelm a single prior factor.  The presence of member-level uncertainties
implies that each new object \emph{adds a new parameter}, so differences
between Bayesian and likelihood estimates persist.  This is evident in
the Malmquist corrections described above.  The only
way for these Bayesian estimates to converge to MLEs is to add follow-up
data for each source, i.e., to make all of the $\ell_i(M_i)$ functions
narrower.

Considering the population parameter $M_0$ to be unknown, with prior
distribution $\pi(M_0)$, the joint posterior for all the unknowns is given by
\begin{equation}
\pi(M_0,\vM|\vD) \propto \pi(M_0) \prod_i f(M_i|M_0)\, \ell_i(M_i).
\label{LH-joint-post}
\end{equation}
If our goal is to estimate the source parameters, we account for $M_0$
uncertainty by marginalizing over $M_0$, giving the source parameter marginal
posterior, $\pi(\vM|\vD) = \int \dif M_0\, \pi(M_0,\vM|\vD)$.  If instead our goal is
to infer the population density, we calculate the marginal posterior for $M_0$,
$\pi(M_0|\vD) = \int \dif\vM\, \pi(M_0,\vM|\vD)$.  In this simple normal--normal MLM,
these integrals can be done analytically (we adopt a flat prior for $M_0$---the
\emph{hyperprior}).

\begin{figure}[t]
\centerline{\includegraphics[width=.8\textwidth]{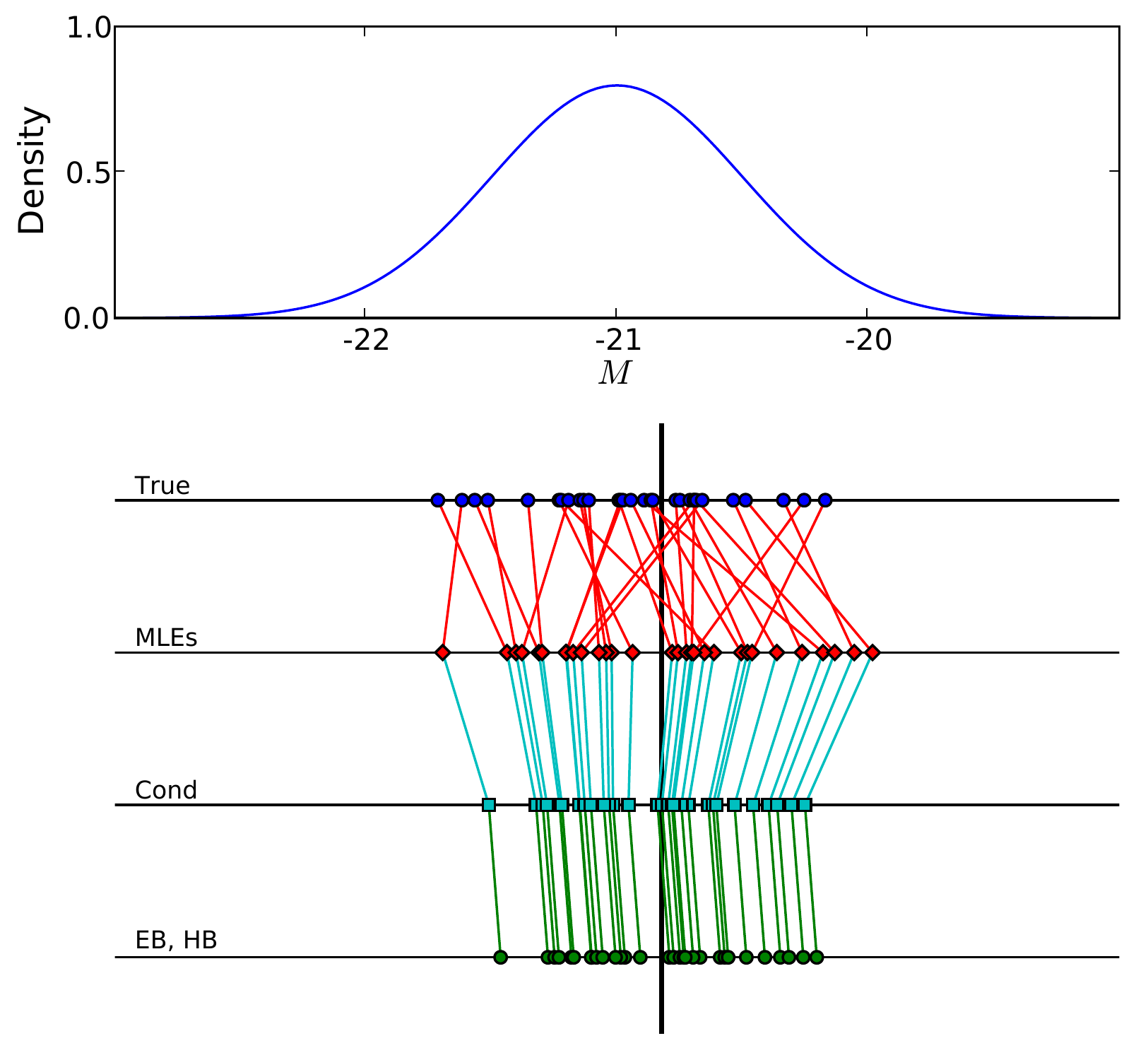}}
\caption{\small Shrinkage in a simple normal--normal model.  
Top panel shows population distribution; ``True'' axis shows
$M_i$ values of 30 samples.  Remaining axes show estimates from
measurements with $\sigma=0.3$ normal error: MLEs, conditional
(on the true mean), and empirical/hierarchical Bayes estimates.}
\label{LH-GaussShrink}
\end{figure}

The top panel of figure~\ref{LH-GaussShrink}\ shows the PDF for a population distribution
with $M_0=-21$; the circles on the line just below it indicate the true $M_i$
values of a sample of $N=30$ objects.  Below that, diamonds indicate the
MLEs for the objects, $\hat M_i$, for one realization of measurement
error; a line segment connects each estimate with its true parent value.
The MLEs are intuitively appealing estimates; they also have several appealing 
frequentist properties, considering an ensemble of many realizations
of the measurement errors for a particular object.  For example, normal MLEs are unbiased (in
fact, they are the best linear unbiased estimators),
and they are invariant to translation in $M$.  But considered as an
ensemble, they are overdispersed with respect to the population distribution;
this is visually evident in the figure.
Intuitively, the MLEs here can be viewed as samples from a modified population density
that is the convolution of the true population density with the error density
(though we note that this convolution interpretation does not generalize to more
complicated settings).

As alternatives to the MLEs, consider Bayesian estimates, in two different
scenarios.  First, suppose we knew $M_0=-21$ a priori.  Using the (known)
population distribution as the prior for each $M_i$ produces posteriors that
remain independent and normal, but with means, $\tilde M_i$, shifted from the
MLEs toward $M_0$.  Define $b \equiv \frac{\sigma^2}{\sigma^2 + \tau^2}$; then
the posterior means (and modes) are given by
\begin{equation}
\tilde M_i = (1-b)\cdot\hat M_i + b\cdot M_0,
\label{LH-M-M0}
\end{equation}
and the variance for each estimate is reduced to $(1-b)\sigma^2$
rather than just $\sigma^2$.  The squares on the third line below
the panel in figure~\ref{LH-GaussShrink}\ show these estimates. They
all move toward $M_0$, and thus toward each other. One says that the
ensemble of estimates ``shrinks toward $M_0$;'' this phenomenon is
called \emph{shrinkage}.  They are labeled ``Cond'' in the figure to
indicate that we conditioned on $M_0$.

As an ensemble, the shrunken estimates look more like the true
values than the MLEs.  These estimates are biased and are no longer
invariant, but even from a frequentist perspective they may be
deemed better than the MLEs: despite the bias, the shrunken
estimates are, on the average (over error realizations, and across the population), closer to
the true values than the MLEs---i.e., they have smaller mean
squared error (MSE)---as long as $N>2$.  Stein discovered this
effect around 1960, and after a decade or two of sorting out its
subtleties, the use of deliberately (and carefully) biased
estimators for joint estimation of related quantities is now
widespread in statistics (frequentist and Bayesian), and considered
one of the key innovations of late-20th century statistics.

We have used strong prior information here---precise knowledge of $M_0$.
But what if we did not know $M_0$ a priori?  The
{\em empirical Bayes} (EB) approach ``plugs in'' an ad hoc estimate of $M_0$
and uses the resulting prior.  The obvious estimator here is $\bar M$, the
average of the MLEs, whose position is indicated by the thick vertical line in
the figure.  Using the resulting prior produces the circle estimates on the bottom
axis, still shrunken, but towards $\bar M$ rather than $M_0$.
Equation~\eqref{LH-M-M0}\ again gives the estimates, if we replace $M_0$ with $\bar M$.
Note that since $\bar M$ depends on {\em all} of the MLEs, the EB estimates
are \emph{no longer independent}.

From a fully Bayesian point of view, plugging in $\bar M$ for $M_0$ is unjustified;
$M_0$ is unknown, so we should consider it a parameter, assign its prior
distribution, and marginalize over it, an approach called {\em hierarchical
Bayes} (HB).  The resulting estimates are identical to the EB estimates in
this problem (though they need not be in general); however, the uncertainties
in the HB estimates are somewhat larger than those produced in an EB
calculation, reflecting uncertainty in the shrinkage point.  We have motivated
EB and HB shrinkage estimates via Bayesian arguments, but the frequentist
advantages of shrinkage in the conditional case still hold:  despite their
bias, as a group these estimates have smaller MSE than the MLEs.  In addition, due to their
accounting for $M_0$ uncertainty, confidence intervals based on the HB
procedure have more accurate frequentist coverage than EB intervals (Carlin and Louis 2000).

The conditional shrinkage estimates are the normal--normal model counterparts
to homogeneous Malmquist and Lutz-Kelker corrected estimates (the latter
do not so obviously ``shrink'' because the prior in astronomical settings
is much broader than
a normal distribution).  On the one hand, this is a source of comfort, in
that some of the statistical benefits of shrinkage are presumably
shared by the astronomical methods.  However, the correspondence is also
a source of caution and concern.  Decades of study have revealed shrinkage
to be a subtle phenomenon, with snares for the unwary.  We highlight just
a few key developments here; entries to the large literature in this
area include Carlin and Louis (2000), Carlin et al.\ (2006), and
Browne and Draper (2006).

Most obviously, as noted in Section~\ref{LH-sec:estim},
the homogeneous density assumption of
the classic corrections is seldom justifiable; in reality, the population
density is inhomogeneous---and unknown.
That is, conditional shrinkage is not appropriate;
something along the lines of the EB or HB approaches is in order.  This
must be done with some care: it is known that shrinkage may not improve
estimates, and may even worsen them, if done in a matter that does not
reflect the true distribution of lower level parameters.  The EB
approach---using a ``plug-in'' estimate of hyperparameters to specify the
population hyperprior---has appealing simplicity;
the Landy-Szalay approach probably has an approximate EB justification.  But
it is known that EB approaches tend to underestimate final uncertainties
(due to ignoring hyperparameter uncertainty).  HB estimates can offer
improvements, but with computational costs, and with other challenges noted below.

More subtly, shrinkage must be tuned, not only to the underlying
population distribution, but also to the \emph{inferential goal}.  The
shrinkage estimates just described do indeed reduce the MSE of the
collection of source absolute magnitudes.  But if one uses the
shrunken point estimates to infer the population distribution,
it turns out the point estimates are \emph{under}dispersed and the distribution
may be poorly estimated.  If one instead seeks
from the beginning point estimates that are optimal for estimating the
population distribution (via a decision-theoretic calculation), a different
shrinkage prescription is appropriate.  Similarly, it is evident from
the segments connecting the true $M_i$'s with their MLEs that the
ranks of the sources are shuffled, and shrinkage has not corrected it.
In some settings, shrinkage estimates that improve rank estimates
have been identified; they differ from those optimal for individual
parameter or distribution estimates.  There is a kind of ``complementarity''
in relying on point estimates for subsequent inferences; estimates
optimal for some questions may be misleading for other questions (Louis 1984).

A main source of these complications is the inadequacy of point estimates as
summaries of a correlated, high-dimensional posterior distribution.  This
motivates a more thoroughly Bayesian treatment in the spirit of HB, relying on
marginalization over uncertain parameters rather than use of point estimates.
We pursue this approach below.


But before doing so, some
comments at a conceptual level are appropriate here.  The second
level of our MLM here describes the population with a continuous
density.  A frequentist interpretation of this density is
problematic.  The volume accessible to a survey---indeed, the volume
within the horizon---contains a finite number of sources (galaxies,
clusters, quasars, gamma-ray bursts, etc.).  Repeating a
survey will produce catalogs that largely contain the same sources
(some sources near the survey detection limit may differ from one
repetition to the next); the population density cannot be
interpreted in terms of frequentist variability.  At the lower level
of the MLM, measurement errors may differ among repetitions, but if
the lower level uncertainties are the result of indicator
scatter---itself a population-level phenomenon---lower-level results
will also be the same across repeated surveys.

From the Bayesian point of view, the population PDF describes a priori \emph{uncertainty}
about the value of a property for a population member, not (directly) \emph{variability} in repeated sampling; its introduction
and specification should
be motivated by epistemological considerations.  One way
to formally motivate it is as a mechanism to introduce dependence among
estimates.  That is, we expect that learning the properties of many
objects of a particular type should help us predict the properties of
as-yet unmeasured objects of that type; this is what it means to consider
the objects to comprise a population.  Consistency requires that, once
we obtain measurements of new objects, we cannot ignore the prior information
provided by measurements of other objects that we would have used in the
absence of the data.  The resulting dependences in the joint posterior
pools information, leading to shrinkage.

More formally still, we might justify introducing a population density by
requiring the joint prior distribution for a set of object properties
to be {\em exchangeable}, that is, invariant to permuting the labels
of objects.  E.g., in the setting above, $p(M_1, M_2, \ldots, M_N|\paramvec)$
should take the same functional form if we permute the order of the $\{M_i\}$.
This appears to be both a natural and a weak assumption.
It is in the spirit of the common ``IID'' (independent and identically distributed) assumption in that the marginals
for each source are identical; but by allowing dependence, it sets the
stage for sharing of information across the population.  
Surprisingly, \emph{exchangeability
itself implies the hierarchical Bayes structure}:  the (continuous) de~Finetti exchangeability theorem
implies that any such exchangeable distribution can be written as a
density-weighted mixture of identical, conditionally independent distributions,
i.e., as a hierarchical Bayesian model.  In the setting here, the theorem says one may write
\begin{equation}
p(M_1, M_2, \ldots, M_N) =
  \int \dif\paramvec\; \mu(\paramvec) \, \prod_i f(M_i|\paramvec),
\label{LH-exch}
\end{equation}
where $\mu(\paramvec)$ defines a unit-normed measure over $\paramvec$, specifying the form
of the independent densities $f(\cdot|\paramvec)$.  The theorem is a
purely mathematical result providing a representation for symmetric
functions, but in a Bayesian context it motivates introducing a
continuous population density, playing the role of $f(\cdot)$, with
a hyperprior playing the role of $\mu$.\footnote{Rigorously, the
theorem requires that the judgement of exchangeability apply for any
selection of a finite set of $M_i$'s from an infinite set.  If there
is a finite limit to $N$, the integral representation may not be able to
represent some possible exchangeable distributions, though the
restriction is minor if the limit is large.  See Diaconis and
Freedman (1980) for details.}



\subsection{Thinned latent marked point process models}
\label{sec:TLaMPP}

While the MLM setup above has the essential ingredients needed for us to
move beyond point estimate-based population modeling, we need to generalize
it in two ways to meet the needs of astronomical survey analysis.  First,
the analysis above took the catalog size, $N$, as given.  In an astronomical
survey, $N$ is instead determined by the population density and the
volume surveyed; it is thus informative about the population.  Second,
astronomical surveys suffer from selection effects, most typically in
the form of thinning or truncation in a ``blind,'' ``ab initio'', or ``blanket'' survey (where sources may be
missed due to detection criteria), or censoring in a targeted follow-up
survey (where sources known to exist may have unmeasureable properties
due to limited sensitivity).  We discuss the thinning/truncation case here.

To allow catalog size to be informative about the population
density, we model the population with an inhomogeneous (marked) Poisson point
process, characterised by an \emph{intensity function} rather than a
probability density function.  
The Appendix develops this model in detail; here we outline its main features.
Let $\oparam$ denote the per-object latent parameters specifying properties of interest for an object (e.g., flux, redshift, luminosity, size, morphology, etc.).
The Poisson point process assumption implies
there is an intensity function, $\intens(\oparam)$ that, when known,
allows us to write the probability for there being an object with
$\oparam$ in the interval $[\oparam,\oparam+\dif\oparam]$ as
$\intens(\oparam)\dif\oparam$, to leading order in $\oparam$.  It also
presumes that this probability is independent of whether an object is
found in any other (distinct) interval (provided we know the
intensity $\intens(\cdot)$; i.e., this is \emph{conditional}
independence). Usually we will not know the intensity, e.g., it may
depend on parameters, $\paramvec$, whose values are uncertain, which
we indicate by writing $\intens(\oparam;\paramvec)$.

To account for (random) truncation, we introduce a survey detection
efficiency, $\effic(\oparam)$, specifying the probability that an
object with parameters $\oparam$ will be detected.  Although we take
$\effic(\oparam)$ as given in what follows, it is worth noting
(especially for non-astronomer readers) that calculating the
$\effic$ which characterises a particular survey is often a very
difficult task, requiring both careful measurement and calibration,
and often extensive Monte Carlo simulation.  Not all surveys provide
an accurate detection efficiency, yet it is necessary for what
follows. (In some cases it may be possible to partially infer
$\effic$ from the available survey data; we do not cover this
somewhat subtle task here.)

Finally, we highlight a point made in passing above:  from the
Bayesian point of view, a survey catalog should not be viewed
as providing \emph{estimates} of object properties, but rather as providing
summary statistics specifying \emph{member likelihood functions},
$\ell_i(\oparam_i) = p(D_i|\oparam_i)$, where $D_i$ denotes the data
for object $i$. This change in viewpoint has far-reaching
implications.  It can enable more accurate accounting of object
uncertainties, e.g., by reporting a likelihood parameterisation more
complex than the traditional ``best-fit $\pm$ uncertainty,'' such as
a parameterisation describing possible likelihood skewness (as might
be important near detection limits). It also opens the door to use
of marginal detections or upper limits in censored surveys, by
averaging over object uncertainties with likelihoods that are significantly non-Gaussian, possibly peaking at zero object flux.

With these ingredients in hand, one can calculate the thinned latent Poisson
point process counterpart to the MLM joint posterior density for the population
and source parameters of equation~\eqref{LH-joint-post} (see Loredo \& Wasserman 1995 or Loredo 2004 for early derivations; a more thorough and general treatment is in the Appendix):
\begin{eqnarray}
\pi(\paramvec,\{\oparam_i\}|D) \propto \pi(\paramvec)
\exp\left[-\int \dif\oparam\, \effic(\oparam)\intens(\oparam;\paramvec)\right]
\prod_{i=1}^N  \ell_i(\oparam_i)\intens(\oparam_i;\paramvec).
\label{LH-pp-post}
\end{eqnarray}
Marginal posteriors for $\paramvec$ or $\{\oparam_i\}$ may be
calculated as was done above, though obviously the calculations can
be challenging for astrophysically interesting models.  Our own
applications to date have been to situations with parametric
population models, where $\oparam$ was one-dimensional (magnitudes
of trans-Neptunian objects (TNOs); Petit et al. (2007) and
references therein), or three-dimensional (fluxes and directions of
gamma-ray bursts; Loredo \& Wasserman 1998a,b). In these cases the
$N$ integrals over $\oparam_i$ were done by quadrature.

\begin{figure}[t]
\centerline{\includegraphics[width=.98\textwidth]{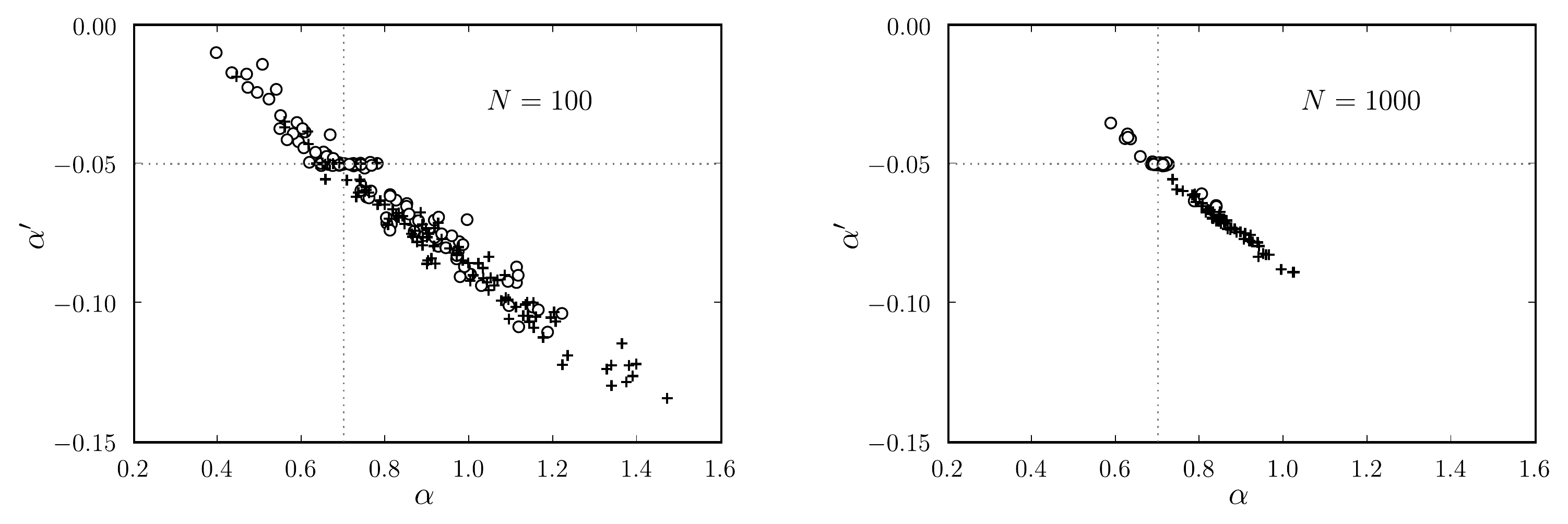}}
\caption{\small Scatter plots of posterior modes for population parameters,
using data simulated from a rolling power law, from a Bayesian analysis
marginalizing over source parameters (circles), and a maximum
likelihood analysis using best-fit source estimates ($+$ symbols).  Left panel
is for simulated surveys of $N=100$ sources; right panel is for $N=1000$.}
\label{LH-scatter}
\end{figure}

As a simple example focusing on one aspect of the MLM approach---the
value of marginalizing over source uncertainty---consider a
magnitude survey (i.e., the ``number counts'' setting), where
$\oparam = m$, the apparent magnitude of a source.  Suppose the
source population has a rolling power law distribution of fluxes, so
we may write the magnitude distribution as $\intens(m) = A\times
10^{[\alpha(m-23) + \alpha'(m-23)^2]}$, where $A$ is the density per
unit magnitude at $m=23$, and $\alpha$ and $\alpha'$ give the slope
of the number-magnitude distribution, and its rate of change with
$m$, at $m=23$.  We simulated sources from this distribution, and
simulated detections and measurements for a very simple survey
performing source detection and measurement via photon counting. The
survey parameters were chosen so that the dimmest detected sources
have magnitude uncertainties $\sim 0.15$ magnitudes.  We analysed data from
many simulated surveys, all of a population with $\alpha=0.7$ and
$\alpha'=-0.05$ (values that describe some TNO data), and estimated
$\alpha$ and $\alpha'$ by finding the mode of the marginal posterior
for these parameters, marginalizing equation~\eqref{LH-pp-post}\ over
all the latent $m_i$ parameter, and the amplitude, $A$.  The resulting $(\alpha,\alpha')$ estimates, for
surveys of size $N=100$, are plotted as the open circles in the left
panel of figure~\ref{LH-scatter}; the crosshair shows the true
parameter values.  We also calculated maximum likelihood estimates of $\alpha$ and $\alpha'$, maximizing the full joint likelihood also over the latent parameters,
i.e., using only the best-fit $m_i$ estimates (not marginalizing).
These are plotted as ``$+$'' symbols in the figure.  The Bayesian
estimates are distributed roughly symmetrically about the truth; the
MLEs are clearly biased toward large $\alpha$ and small $\alpha'$,
though the uncertainties are large enough that the estimates are
sometimes accurate.

The right panel shows results from the same calculation, but with $N=1000$.
The Bayesian estimates have converged closer to the truth.  In contrast, the
MLEs have converged \emph{away} from the truth.  This example highlights the
value of marginalization, particular in settings with measurement error.  In
such settings, each new object adds its member parameter(s) to the problem; one is not in
the fixed parameter dimension setting in which our statistical intuitions about ``root-$N$ convergence'' are
trained.  As a result, the effects of object uncertainties do not ``average
out;'' instead, it becomes \emph{more} rather than less important to carefully
account for them as survey size grows.  This aspect of population modeling has
been repeatedly rediscovered by astronomers.  The earliest discovery we know
of is Eddington's treatment of what has become known as \emph{Eddington bias} (Eddington 1940).
A recent rediscovery in a cosmological context is 
Sheth's work on the effects of photometric redshift errors on modeling galaxy
and quasar populations (Sheth 2007).  Sheth advocates an ad hoc deconvolution
algorithm; we think hierarchical Bayes offers a vastly more flexible
and accurate framework for addressing such problems.


\section{Research directions}
\label{LH-sec:future}

Fully Bayesian hierarchical modeling of cosmic populations is challenging.
Most applications to date either rely on fairly simple models, or, when considering complex models, consider small or modest-sized surveys.
The earliest HBM work in cosmic demographics appears to be the work of Loredo \& Wasserman (1995, 1998a,b) on gamma-ray burst (GRB) demographics; they adopt parametric number-counts and luminosity function models for samples of $\sim 10^3$ GRBs.
The challenging VELMOD analysis of Tully-Fisher data 
by Willick and Strauss 1998, mentioned in Section~\ref{LH-sec:app}, includes
many key elements of hierarchical Bayes, but does not implement a fully Bayesian calculation.
The state-of-the-art for large-scale multilevel modeling of survey data is probably the reanalysis of the Sloan Digital Sky Survey (SDSS) catalog by Regier et al.\ (2019), but this calculation adopts fixed population distributions, and invokes strong approximations to enable large-scale computations using a variational inference algorithm.
See Loredo (2013) for a recent survey of hierarchical Bayesian modeling of cosmic populations.
Future research must explore applications of
increased complexity in three different dimensions:  survey size, source
parameter dimension, and population model complexity.

An obvious need is development of numerical algorithms appropriate
for multivariate observables and large surveys, perhaps invoking
approximations or using Monte Carlo methods and parallel computing for member latent parameter
marginalization (see Szalai-Gindl etal.\ 2018 for promising work along these lines using GPUs to implement parametric HBMs for population sizes $\sim 10^6$, with low-dimensional member properties).
With respect to multivariate observables, it will
be insightful to work out the detailed connections between the MLM
approach and other methods, such as those surveyed in
Section~\ref{LH-sec:indic}.  For example, when the observables are
flux and a distance indicator, an analogue to the direct indicator
method should ``fall out'' of an MLM calculation when the source
likelihoods provide precise estimates of the indicators.

Implementing Bayesian MLM with more complex population models will also
demand development of clever algorithms.  But more subtle and interesting
challenges arise as model complexity increases.  These challenges arise
because of the ``softening'' of the impact of member-level measurements on inferences,
due to uncertainties.  We saw above that this ``changes the rules'' in
the sense of causing violations of naive intuition about uncertainties
averaging out as $N$ grows.  But this is only one of several issues
complicating life with multilevel models.

Some of these issues mimic problems associated with nonparametric modeling,
and this is no accident.  Though a common informal ``definition'' of
nonparametric model is a model with an infinite number of parameters, a more
insightful definition is a model in which the \emph{effective} dimension of
the parameter space can grow with sample size.  In fact, the ``many normals''
problem---essentially the normal--normal MLM, with the actual dimension
growing linearly with sample size---is sometimes used as a
surrogate for more complex nonparametric models in theoretical analyses
(e.g., Wasserman 2005).

A prime issue which nonparametric modellers must face is assessing
how priors over large-dimensional spaces may influence inferences.
Similar concerns arise for MLMs.  For example, for estimating the
mean and standard deviation of a normal distribution using
\emph{precise} measurements, common default priors are flat for the
mean and log-flat for the standard deviation.  We saw above that a
flat prior for $M_0$ in the normal--normal MLM produced sensible
inferences.  But had we considered the population standard deviation
$\tau$ to be unknown, we would have discovered that a log-flat
$\tau$ prior leads to an improper (unnormalisable) posterior.
Instead, priors flat in $\tau$ or $\tau^2$ (among others) are
advocated, with various justifications (sometimes including the good
frequentist performance of the resulting estimates; see, e.g.,
Berger et al.\ 2005; Gelman 2006). This indicates that as
astronomers increase the complexity of MLMs for surveys, care must
be taken with priors; statisticians have helpful insights to offer
here.  It also suggests that model checking, in the spirit of
goodness-of-fit tests, is important for MLMs.  Their rich structure
makes conventional model checking methods inappropriate, but there
is useful research on model checking methods tailored to MLMs (e.g.,
Sinharay \& Stern 2003; Bayarri \& Castellanos 2007).

An alluring direction for increasing model complexity is to make the
population model itself truly nonparametric.  One motivation comes from
existing nonparametric methods designed to flexibly account for selection
effects, such as the $C^-$ method of Lynden-Bell (1971) or the stepwise maximum
likelihood method (Efstathiou et al.\ 1988).  These methods ignore object uncertainties;
finding counterparts in the MLM framework promises to broaden applicability
of such approaches, and unify them with approaches relying on
Malmquist-style corrections (insofar as MLMs have shrinkage ``built in'').
But the fact that \emph{parametric} MLMs already have some of the
issues of nonparametric modeling suggests that nonparametric multilevel modeling
will be tricky, requiring even more care with assessing robustness to priors.
Fortunately, there are successful examples in the statistics literature
to build upon (e.g., M\"uller \& Quintana 2004).

Sensitivity to priors can make newcomers to Bayesian methods consider
retreating to frequentist territory.  But there is little solace there;
the subtleties of complex multilevel Bayesian modeling reflect genuine
complexity in the task of modeling surveys, complexity that has frequentist
implications.  For example, the best-studied methods for nonparametric
analysis of survey data with uncertainties (yet to be extended to include
truncation) rely on deconvolution.  Though the methods are straightforward,
it is known that the resulting population estimates have discouragingly
slow rates of convergence (often only logarithmic in $N$, as opposed to
the $\sqrt{N}$ we are accustomed to for parametric inference without
measurement errors; see Loredo 2007 for discussion and entries to the literature).  In fact, leading developers of such methods have
recently turned to Bayesian methods, where careful attention to
structure in the prior can lead to methods with improved performance
(e.g., Berry et al.\ 2002).

We think development of \emph{semiparametric} population models for use in MLM
analyses of survey data may be an especially fruitful research direction.
For example, we envision nonparametric modeling of the density
distribution of galaxies, combined with parametric modeling of the luminosity
function (e.g., by a Schechter function or a mixture of a few Schechter
functions), as a promising approach, allowing adaptivity to complex spatial
structure, but hopefully providing good convergence rates for learning the
luminosity function.  But there are several challenges to conquer on the
path from the current state of the art to such a goal.

We are grateful to many collaborators and colleagues who have
contributed to our understanding of survey biases and population
modeling, especially David Chernoff, Woncheol Jang, David Ruppert, John Simmons,
and Ira Wasserman.  We also gratefully acknowledge the Statistical
and Applied Mathematical Sciences Institute (SAMSI), whose 2006
Astrostatistics Program and 2016 ASTRO Program assembled astronomers and statisticians to
discuss these issues.  Loredo was partly supported by NSF grant AST-0507589
and by NASA grants NAG5-12082 and NG06GH84G for work reported here.
Both Loredo and Hendry were supported by NSF grant AST-1312903 for work described in the appendix.

\newcommand{\dd}{\textrm{d}}

\newcommand{\assm}{\,|\,}
\newcommand{\mlike}{\ell}
\newcommand{\ppar}{\theta}  
\newcommand{\obsv}{\mathcal{O}}  
\newcommand{\psivec}{\boldsymbol{\psi}}
\newcommand{\Dvec}{\boldsymbol{D}}
\newcommand{\lfunc}{\phi}  
\newcommand{\clfunc}{\Phi}
\newcommand{\lpdf}{f}
\newcommand{\lcdf}{F}
\newcommand{\rhopar}{\boldsymbol{\zeta}}
\newcommand{\dtxns}{\mathcal{D}}
\newcommand{\Fth}{F_{\rm th}}
\newcommand{\Fhat}{\hat{F}}
\newcommand{\Ffid}{F_{\rm fid}}
\newcommand{\rmax}{r_{\rm max}}
\newcommand{\aeffic}{\overline{\eta}}
\newcommand{\epdf}{\mu}
\newcommand{\Lsol}{L_\odot}

\section*{Appendix: Thinned latent marked point process\\ likelihood function}
\renewcommand{\theequation}{A.\arabic{equation}}
\renewcommand{\thefigure}{A.\arabic{figure}}




In this Appendix we derive the likelihood function for the parameters of a population model and the population's latent parameters, presented as equation~\eqref{LH-pp-post} in \S~\ref{sec:TLaMPP}.
As noted there, this likelihood function is based on catalog data describing member likelihood functions for detected objects, and appropriate summaries of the detection criteria to account for selection effects; here we describes those features in more detail.
For concreteness, we treat the case of using galaxy survey data to infer a luminosity function.
The latent object parameters are flux, direction (or 2D position on the detector), and distance, $\oparam_i = (F_i, x_i, z_i)$.
For simplicity, we treat the case where the data provide noisy flux estimates, but where redshift is precisely known (e.g., considering a sample that supplements photometry with high-resolution spectroscopy with good signal-to-noise).

Object detection is typically implemented via a scanning procedure.
For example, for image data, a fixed aperture may be scanned over the image.
As the scan proceeds, a detection algorithm determines if an object is present at each candidate location, e.g., by comparing the estimated flux in the aperture to a threshold value (set by background and noise estimates), or by fitting an image model to the data in the aperture and comparing the fitted amplitude to a threshold.
For time series data, a window may be scanned over the time series, with an object detected if the estimated flux in the window is above a threshold.
If an object is detected, its properties are more carefully estimated, e.g., by a likelihood-based or weighted least-squares calculation, with estimation results summarized in the catalog.

\begin{figure}
\begin{center}
\includegraphics[width=.8\textwidth]{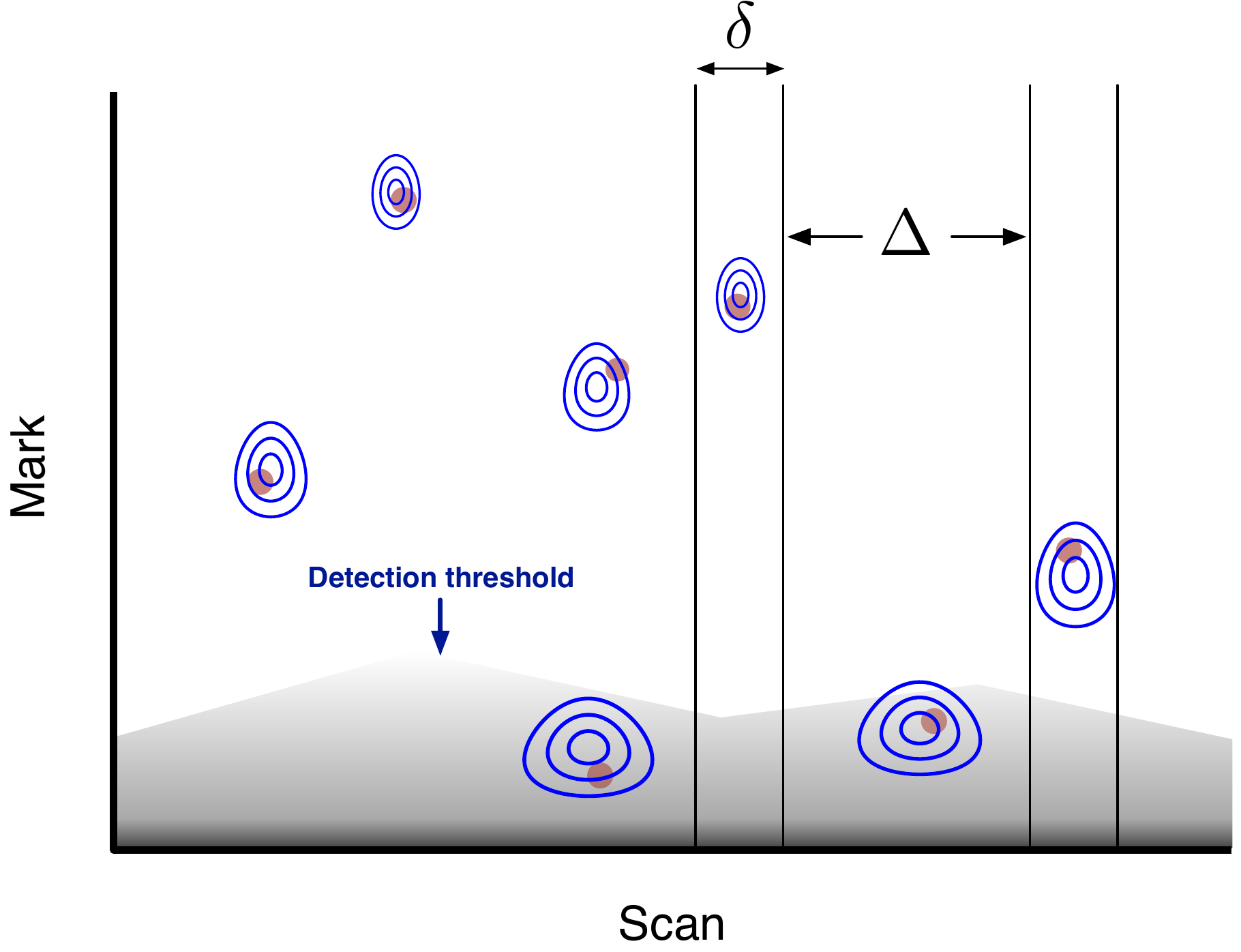}
\end{center}
\caption{Depiction of thinned latent marked point process model for catalog data produced by an astronomical survey.
Object properties are split into a scanned subset and a mark subset.
Dots (red) show latent (true) values for an object's properties.
Contours (blue) depict member likelihood functions from analysis of the raw survey data; catalogs provide summaries of these for detected objects.
Gray region at bottom depicts the non-detection region; candidates with estimated mark values below a varying threshold are rejected.
$\delta$ and $\Delta$ denote sizes of example detection and nondetection intervals.}
\label{fig:ScanMark}
\end{figure}

Fig.~\ref{fig:ScanMark} illustrates the process and its relationship to catalog construction.
We split the object property parameter space into \emph{scan} and \emph{mark} components.
The scan component corresponds to the dimensions over which the detection scan operates; 
the mark component corresponds to the remaining dimensions.
For our galaxy luminosity function example, the scan component is the two-dimensional position of the galaxy image on the detector (corresponding to its direction on the sky), and the mark component is the galaxy luminosity and distance, or equivalently, flux and distance (in a more complex case, the mark component might include color and morphological parameters).
In the figure, the dots (red) indicate the true properties of seven galaxies; the blue contours depict likelihood functions for the properties, based on noisy image data (the displayed contours are from simulations using a simple 1-D image model, with a single location parameter, and a flux parameter).
The gray region at the bottom is bounded above by the position-dependent detection threshold; an object is detected only if its best-fit (maximum likelihood) flux is above the threshold.
The region is depicted with a gradient to depict that, as a function of \emph{true} object properties, the resulting selection is probabilistic.
Here two of the seven objects are not detected.

We model the properties using a (latent) marked Poisson point process, i.e., a Poisson point process for the scanned parameters, and a probability density function for the mark parameters.
For concreteness, we focus on the luminosity function example, taking the scan parameter to be object position, $x$ (a 2D parameter, e.g., direction on the sky), and the mark parameters to be flux and distance, $(F, r)$.
We suppose that the spatial density of galaxies is approximately constant over the region probed by the survey.
There is thus a constant Poisson intensity parameter, $\lambda$, specifying the density of galaxies in $x$.
We assume a luminosity PDF that is independent of distance (of course, the flux PDF will depend on distance, thanks to the inverse square law).
The flux and distance mark PDF is thus a product of a distance PDF, $h(r)$, and a conditional flux PDF, $\rho(F,r)$.
We use $\rhopar$ to denote the flux PDF parameters, writing it as $\rho(F,r; \rhopar)$ when we want to display the parameter dependence.

For the PDF for galaxy distance, $h(r)$, we assume homogeneity, which implies
\begin{equation}\label{eq:distPDF}
h(r) = 
\begin{cases} 
    \dfrac{3r^{2}}{r_u^{3}} & \quad \text{if } 0\leq r\leq r_u,\\
    0 & \quad \text{otherwise},
\end{cases} 
\end{equation}
where $r_u$ is an upper limit on distance chosen to be beyond the surveyed volume.
(That is, $r_u$ is chosen so that the most luminous galaxies of interest have fluxes comfortably below the lowest flux threshold.
In deep surveys, reaching to very dim fluxes, cosmological considerations, including the finite age of the universe and the non-Euclidean geometry of spacetime, ameliorate the growth of $h(r)$ with $r$.)
The population model thus has parameters $\ppar = (\lambda,\rhopar)$.

We consider a case where we have precise distance measurements for the galaxies (e.g., from high-resolution spectroscopic data providing precise redshifts).
We assume independent errors in the position and flux measurements, so the catalog contains descriptions of separate member likelihood functions for flux and position, denoted $\ell_i(F)$ and $m_i(x)$ for galaxy $i$, with $i=1$ to $N$.%
\footnote{Independence of flux and position estimates is almost universally assumed for astronomical catalogs, but for dim sources there can be significant dependence.
We depict this in Fig.~\ref{fig:ScanMark}.}
Formally, denoting the image data for detected galaxy $i$ by $D_i$, we are writing
\begin{equation}\label{eq:xFr-like}
p(D_i|x,F,r) = \ell_i(F)\, m_i(x)\, \delta(r - r_i),
\end{equation}
where the Dirac delta function factor represents the precise measurement of distance.

We must also describe the survey's selection effects.
These are determined by the detection threshold as a function of the scan location.
At each scanned location, $x$, the threshold determines the set, $\dtxns_x$, of possible data (i.e., arrangements of counts in the pixels in a scanned aperture) that would pass detection criteria.
For example, if the detection criterion is that the MLE flux estimate, $\hat F(D)$ for data $D$, must exceed a threshold $\Fth(x)$, then $\dtxns_x = \{D : \hat{F}(D) > \Fth(x)\}$.
Reporting $\dtxns_x$, or equivalently $\Fth(x)$, then describes the selection effects.
But we will see below that a more compact summary of the detection criteria will be more convenient.

We now compute the likelihood function for the parameters, based on catalog data describing member likelihood functions and the selection effects.
For simplicity, we here consider ``nearly-pure catalog'' settings with stringent detection criteria (e.g., high thresholds), so that it is unlikely there are any false detections in the catalog (it is straightforward to generalize to settings with nonnegligible false detection rate).
Fig.~\ref{fig:ScanMark} includes depictions of elements of our construction.
We partition the scan space into $N$ detection intervals, $\delta_i$, containing a single detected object, and $M$ nondetection intervals, $\Delta_j$, in which no candidate object passed the detection criterion.%
\footnote{We are presuming that galaxy images are well-separated, i.e., we do not treat here the \emph{crowded field} or \emph{strongly blended} case, where the images of distinct objects may strongly overlap.}
The likelihood function is the product of the (conditionally independent) probabilities for these intervals.

We first consider the probability for no detection in one of the $\Delta_j$ intervals.
We break it up into subintervals of size $\delta x$, small enough that the detection threshold is approximately constant over the interval.
The probability for seeing no detections in $\delta x$ is the sum of the probabilities for the following events (conditioned on the population parameters, $(\lambda,\rhopar)$):
\begin{itemize}
\item No objects have $x$ in the interval.
\item One object has $x$ in the interval, but it produced data that were not in $\dtxns_x$.
\item Two objects have $x$ in the interval, but both produced data that were not in $\dtxns_x$.
\item And so on\ldots.
\end{itemize}
Each event is a conjunction of two simpler events, the Poisson probability for the specified number of objects lying in the interval, and the probability for not detecting any events in the interval.
We will express the latter probability in terms of the \emph{detection efficiency} at $x$ for objects with flux $F$,
\begin{align}\label{eq:eta-def}
\effic(x,F) 
  &\equiv p(D\in\dtxns_x | F)\\
  &= p(\hat{F}(D) > \Fth(x) | F),
\end{align}
where $F$ as a conditioning symbol signifies that an object is present with flux $F$.
The probability for detecting an object with unspecified flux and distance, given the population parameters, is then
\begin{equation}\label{eq:p-x}
p_x(\rhopar) = \int \dd r \int \dd F\,\rho(F,r)\, h(r)\, \effic(x,F).
\end{equation}
The probability for \emph{not} detecting an object with a given location is then $1-p_x(\rhopar)$.

Now let $\nu$ denote the (unknown) number of objects with $x$ in $\delta x$.
Then the probability for no detections in $\delta x$ at $x$ is
\begin{align}\label{eq:q-exp}
q(x) 
  &= \sum_{\nu=0}^\infty \frac{(\lambda\delta x)^\nu}{\nu!} e^{-\lambda\delta x}
        \left[1 - p_x(\rhopar)\right]^\nu\nonumber\\
  &= e^{-\lambda\delta x} \sum_{\nu=0}^\infty \frac{(\lambda\delta x)^\nu \left[1 - p_x(\rhopar)\right]^\nu}{\nu!}
          \nonumber\\
  &= \exp\left[-\lambda\delta x p_x(\rhopar)\right].
\end{align}
This is the probability for no detections in a subinterval of a $\Delta_j$ interval.
The probability for no detections across the entire interval is the product of its subinterval probabilities.
The exponents add, comprising an integral over the $\Delta_j$ intervals, so that the nondetection probability becomes
\begin{equation}\label{eq:q-def}
q(\Delta_j) = \exp\left[-\lambda\int_{\Delta_j}\dd x \int \dd r \int \dd F\,\effic(x,F)\, h(r)\, \rho(F,r)\right].
\end{equation}
This is just the Poisson probability for seeing no events, when the expected number of events is $\lambda$ times the fraction of the population expected to be detected in the interval, given the threshold behavior (encoded in the detection efficiency).

Now consider the probability for the data associated with a detection interval, $\delta_i$; for simplicity, we assume all of these intervals are of the same size, $\delta$, in $x$.
The probability for getting data $D_i$ from detection of an object in $\delta_i$ is the sum of the probabilities for the following events:
\begin{itemize}
\item One object has $x$ in the interval, and was detected producing data $D_i$.
\item Two objects have $x$ in the interval, one of which was detected producing $D_i$, with the other undetected.
\item And so on\ldots.
\end{itemize}
To simplify the calculation, let us stipulate that the detected object has values of $(x,F,r)$ known precisely, i.e., lying in small intervals $(dx, dF, dr)$; at the end, we will account for their uncertainty via marginalization.

The first case is simple; the probability for one object in the interval, having the specified properties, and being detected producing $D_i$, is
\begin{equation}\label{eq:dtxn1}
p_1(\lambda,\rhopar) = 
  (\lambda \delta) e^{-\lambda \delta} 
  \left[\frac{dx}{\delta}\, h(r)dr\, \rho(F,r;\rhopar)dF\right]
  p(D_i\in\dtxns_x, D_i|x,F,r).
\end{equation}
The final probability factor here is for a conjunction; it may be written
\begin{equation}\label{eq:dtxn-joint}
p(D_i\in\dtxns_x, D_i|x,F,r) = p(D_i|x,F,r)\, p(D_i\in\dtxns_x | D_i),
\end{equation}
where we have dropped $(x,F,r)$ from the last factor because the values of the properties are irrelevant for determining detection, once the data are in hand.
Now note that detection is deterministic given the data, i.e., either the data correspond to a candidate passing the detection criteria or not.
But for a detected object, by definition the data passed the criteria, so the last factor is equal to unity.
The first factor we recognize as the member likelihood function, defined in (\ref{eq:xFr-like}).
This completes the computation of $p_1(\lambda,\rhopar)$.

For cases with $\nu > 1$ objects present, we will have a factor like $p_1(\lambda,\rhopar)$ for the detected object, and nondection probabilities for undetected objects of the form of the $[1 - p_x(\rhopar)]$ factor appearing in the $\Delta_j$ probability derived above.
But in addition, we have to account for not knowing which of the $\nu$ objects is detected.
The resulting probability for the case of $\nu$ objects present can be written as follows:
\begin{equation}\label{eq:dtxn-nu}
\begin{split}
p_\nu
  &= \frac{(\lambda\delta)^\nu}{\nu!}  e^{-\lambda\delta}\\
  &\quad \times \left(\frac{dx}{\delta}\, h(r)dr\, \rho(F,r;\rhopar)dF\right) \ell_i(F)\, m_i(x)\, \delta(r - r_i) \\
  &\quad \times \left[1 - p_x(\rhopar)\right]^{\nu-1}\\
  &\quad \times \nu.
\end{split}
\end{equation}
Line by line, the factors are:
\begin{itemize}
\item the Poisson probability for $\nu$ objects being in the interval,
\item the probability for one of them having the given properties and
producing the detection data, $D_i$,
\item the probability for the remaining objects not being detected,
\item a factor of $\nu$ from summing over the possibilities for
which of the $\nu$ objects is detected.
\end{itemize}
To facilitate summing the $p_\nu$ probabilities over $\nu$, we rewrite \eqref{eq:dtxn-nu}, gathering the $\nu$-dependent terms on the second line of the following equation:
\begin{equation}\label{eq:dtxn-nu2}
\begin{split}
p_\nu
  &= (\lambda\delta)\,  e^{-\lambda\delta}
  \left(\frac{dx}{\delta}\, h(r)dr\, \rho(F,r;\rhopar)dF\right) \ell_i(F)\, m_i(x)\, \delta(r - r_i) \\
  &\quad \times \frac{1}{(\nu-1)!}(\lambda\delta)^{\nu-1}
    \left[1 - p_x(\rhopar)\right]^{\nu-1}.
\end{split}
\end{equation}
Upon summing over $\nu \ge 1$, and marginalizing over the uncertain values of $(x,F,r)$, we find that the probability for the detection data in interval $\delta_i$ is
\begin{equation}\label{eq:p-dtxn}
p(D_i|\lambda,\rhopar) =
  q(\delta_i)\, h(r_i)\, (\lambda\delta)\,
  \left[\int_{\delta_i}\frac{\dd x}{\delta}\, m_i(x)\right]
  \left[\int \dd F\, \rho(F,r_i;\rhopar)\, \ell_i(F)\right],
\end{equation}
where $q(\delta_i)$ is an exponential of an integral, the same function appearing in the nondetection probability of (\ref{eq:q-def}).

The likelihood function is the product of detection probabilities (\ref{eq:p-dtxn}) and nondetection probabilities (\ref{eq:q-def}) for all of the $\delta_i$ and $\Delta_j$ intervals.
All of these probabilities share an exponential factor resembling (\ref{eq:q-def}).
In the product, there will be a sum of the integrals in the exponents; this corresponds to a single integral over the entire $x$ domain of the survey, of the form:
\begin{equation}\label{eq:int3}
\lambda\int_\Omega \dd x \int \dd r \int \dd F\,\effic(x,F)\, h(r)\, \rho(F,r;\rhopar),
\end{equation}
where $\Omega$ denotes the full range of positions surveyed (which would be measured in terms of solid angle on the sky).
Note that the only $x$-dependent factor in the integrand is the detection efficiency.
This lets us write the integral in simpler manner.
Introduce the \emph{average detection efficiency},
\begin{equation}\label{eq:aeffic}
\aeffic(F) \equiv \frac{1}{\Omega} \int_\Omega \dd x \,\effic(x,F).
\end{equation}
Using this, (\ref{eq:int3}) can be written as a two-dimensional integral,
\begin{equation}\label{eq:int-aeffic}
(\lambda \Omega) \int \dd r \int \dd F\,\aeffic(F)\, h(r)\, \rho(F,r;\rhopar).
\end{equation}
The factor $(\lambda \Omega)$ is the expected number of objects in the surveyed region, which depends only on the $\lambda$ parameter.
The remaining factor is the fraction of these that are detectable; it depends only on the remaining population parameters, $\rhopar$.

Equation (\ref{eq:int-aeffic}) shows that the average efficiency is a kind of sufficient statistic for the survey's threshold behavior.
Although catalog builders must determine the detection efficiency over the entire range of the survey, they need only report the lower-dimensional average efficiency for analysts.
A common way to compute a survey's detection efficiency is via a Monte Carlo injection study (i.e., injecting simulated objects with known properties into the detection software pipeline).
Such studies implicitly do this averaging calculation via Monte Carlo.

We can now write down the full likelihood function for the luminosity function parameters.
Dropping some factors that do not depend on the parameters, the likelihood function is
\begin{equation}\label{eq:like}
\begin{split}
\like(\lambda, \rhopar)
  &= \lambda^N \exp\left[- (\lambda \Omega) \int \dd r \int \dd F\,
            \aeffic(F)\, h(r)\, \rho(F,r;\rhopar)\right]\\
  &\quad\times \prod_{i=1}^N h(r_i) \int \dd F\, \rho(F,r_i;\rhopar)\, \ell_i(F).
\end{split}
\end{equation}
This likelihood function corresponds to equation~\eqref{LH-pp-post} in the main text, specialized to this luminosity function inference problem.

This likelihood function is reminiscent of that for an inhomogenous Poisson point process, whose likelihood is proportional to a product of intensity function factors, evaluated at the observed points, and an exponential whose negative argument is the integral of the intensity function over the observed domain.
One difference is the integral over the latent observable, $F$, in the product factor; this accounts for measurement error.
A more subtle but important difference is that the integrand in the exponential is not the same function playing the role of the intensity function in the product factor.
There is an average efficiency factor in the exponential, but \emph{not} in the product factor.
This is because of a common feature of astronomical surveys noted earlier: the data used for characterization (estimating member latent parameters) are also used for detection.
As a result, were one to insert an efficiency factor into the product terms, some of the data would be doubly used.
Such considerations appeared explicitly in our derivation; see the text after (\ref{eq:dtxn-joint}).
Some heuristic derivations of similar likelihood functions in the astronomical literature have missed this point, instead inserting an $\aeffic(F)$ factor in the detected object integrals in the likelihood function.
This corrupts inferences; see Loredo (2004) for further discussion.

This point deserves some elaboration.
The statistical literature on sample surveys has of course long been concerned with selection bias.
But the typical setting in the sample survey literature is one where an individual is selected by some process, and subsequently fills out a survey (or is otherwise subject to measurement)---\emph{selection is independent of measurement}.
Key concepts in sample survey statistics rely on such independence; an example is \emph{inverse probability weighting}, as appears in the well-known Horvitz-Thompson estimator (closely related to the $V/V_{\rm max}$ method popular in astronomy; Loredo \& Wasserman (1995) discuss relationships between such methods and the framework described here).
But in astronomical surveys measurement and selection (detection) are typically closely linked and are not independent---selection typically relies on measurement (perhaps done approximately), and object characterization relies on data used for detection.
The analysis above treats such cases.
In some cases, part of the selection process may be independent of object characterization.
For example, in high-energy astrophysics and particle astrophysics settings, there may be an anti-coincidence detector, separate from the main detector, used to exclude events that are likely due to uninteresting backgrounds (but also excluding some interesting events).
If the anti-coincidence data are ignored for characterizing detected events, then $\rho(\oparam)$ above should be replaced by
$\tilde\rho(\oparam) = F(\oparam)\rho(\oparam)$, where $F(\oparam)$ is a \emph{selection filter function} that accounts for the independent selection process.
Other settings with this structure are those where, in some sense, nature is selecting a subsample; that is, we have specified $\rho(\oparam)$ as describing some underlying population, not the actually observable population, which is a subsample due to geometric or other effects beyond the control of the observer.
An example in exoplanet demographics arises in modeling transit survey data.
Only planetary systems with the right geometry are observable; if $\rho(\oparam)$ describes \emph{all} systems, then a filter function needs to be introduced accounting for geometry-based selection.
Of course, transit data are subject to other selection effects that are tied to the transit measurement process (e.g., only transits with estimated depth greater than a threshold are deemed detections); in these settings, the presence of a filter function does not imply the absence of a detection efficiency, and both types of selection must be accounted for.

Notably, the Poisson process intensity parameter, $\lambda$, appears in the likelihood function only in two places: in the factor in front, $\lambda^N$, and multiplying the integral in the exponential.
As a result, if we adopt a conjugate prior for $\lambda$ (a gamma distribution), we can easily compute the marginal likelihood function for the $\rhopar$ parameters.
For simplicity, we adopt the limiting case of a uniform prior for $\lambda$.
Marginalizing over $\lambda$ and dropping some $\rhopar$-independent terms, we find that the marginal likelihood function for $\rhopar$ takes the form
\begin{equation}\label{eq:mlike}
\like_m(\rhopar)
  = \prod_{i=1}^N \int \dd F\, \epdf(F,r_i;\rhopar)\, \ell_i(F),
\end{equation}
where we have introduced an \emph{effective pseudo-density} for the latent observables, $F$ and $r$,
\begin{equation}\label{aeq:epdf}
\epdf(F,r;\rhopar) \equiv
  \frac{h(r) \rho(F,r;\rhopar)}
    {\int \dd r \int \dd F\,\aeffic(F)\, h(r)\, \rho(F,r;\rhopar)}.
\end{equation}
Equation~(\ref{eq:mlike}) resembles the familiar likelihood function for a binomial point process (i.e., the likelihood function for a fixed-size sample of points independently distributed in some space, found simply by multiplying the point densities), generalized to account for measurement errors described by the member likelihood functions.
But the analogy is not exact, because the effective pseudo-density is not a PDF for the \emph{latent member parameters} $(F,r)$ (it does not integrate to unity over $(F,r)$); rather, it is a probability distribution for the \emph{data} (up to proportionality).

\begin{figure}
\begin{center}
\includegraphics[width=.8\textwidth]{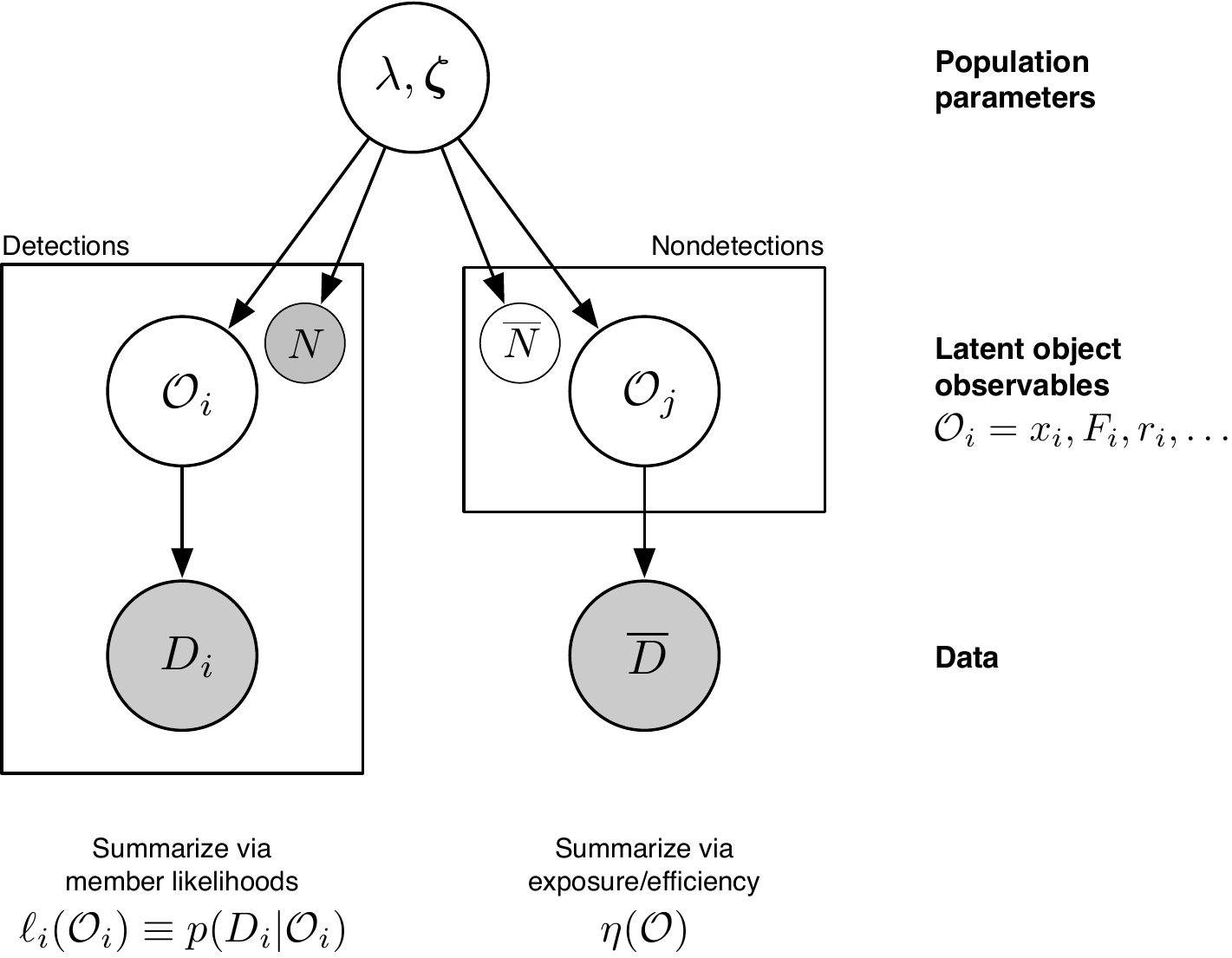}
\end{center}
\caption{Schematic DAG for a thinned latent marked point process (TLaMPP) model for luminosity function estimation from survey catalog data.
The small $N$ and $\overline{N}$ nodes specify the numbers of replications of the detection and nondetection plates, respectively.}
\label{fig:DAG-TLPP}
\end{figure}

Hierarchical and multilevel models are special cases of \emph{probabilistic graphical models}, and their structures are often described using directed acyclic graphs (DAGs)---nodes, denoting a priori uncertain quantities (random variables), connected by arrows indicating conditional dependence (and, importantly, absent edges indicating conditional independence); shaded nodes indicate quantities that become known (data).
Fig.~\ref{fig:DAG-TLPP} shows a schematic DAG for the thinned latent marked point process (TLaMPP) framework.
Separate plates (boxes containing replicated substructures) depict the conditional independence structure for parts of the joint distribution describing detected and undetected objects (a more detailed DAG would partition the nondetection data among the $\Delta_j$ intervals; this would involve nested plates).
The numbers of replications for the detection and nondetection plates, $N$ and $\overline{N}$, are random variables, since the number of objects in the surveyed region is not known a priori, and is informative about the population parameters.

\end{document}